\documentclass[prb,twocolumn,showkeys,showpacs,preprintnumbers,amsmath,amssymb,floatfix,superscriptaddress]{revtex4}

\usepackage{graphicx}
\usepackage{dcolumn}
\usepackage{bm}
\usepackage{dsfont}

\renewcommand{\vec}[1]{\boldsymbol{#1}}
\renewcommand{\Im}{\mathrm{Im}}
\renewcommand{\Re}{\mathrm{Re}}

\begin{document}

\title{Surface Acoustic Wave-Driven Ferromagnetic Resonance in Nickel Thin Films: Theory and Experiment}
\author{L. Dreher}
\email{dreher@wsi.tum.de} 
\affiliation{Walter Schottky Institut, Technische Universit\"at M\"unchen, Am Coulombwall 4, 85748 Garching, Germany}
\author{M. Weiler}
\affiliation{Walther-Mei\ss ner-Institut, Bayerische Akademie der Wissenschaften, Walther-Mei\ss ner-Strasse 8, 85748 Garching, Germany}
\author{M. Pernpeintner}
\affiliation{Walther-Mei\ss ner-Institut, Bayerische Akademie der Wissenschaften, Walther-Mei\ss ner-Strasse 8, 85748 Garching, Germany}
\author{H. Huebl}
\affiliation{Walther-Mei\ss ner-Institut, Bayerische Akademie der Wissenschaften, Walther-Mei\ss ner-Strasse 8, 85748 Garching, Germany}
\author{R. Gross}
\affiliation{Walther-Mei\ss ner-Institut, Bayerische Akademie der Wissenschaften, Walther-Mei\ss ner-Strasse 8, 85748 Garching, Germany}
\author{M.S. Brandt}
\affiliation{Walter Schottky Institut, Technische Universit\"at M\"unchen, Am Coulombwall 4, 85748 Garching, Germany}
\author{S.T.B. Goennenwein}
\affiliation{Walther-Mei\ss ner-Institut, Bayerische Akademie der Wissenschaften, Walther-Mei\ss ner-Strasse 8, 85748 Garching, Germany}

\date{\today}

\begin{abstract}
We present an extensive experimental and theoretical study of surface acoustic wave-driven ferromagnetic resonance. In a first modeling approach based on the Landau-Lifshitz-Gilbert equation, we derive expressions for the magnetization dynamics upon magnetoelastic driving that are used to calculate the absorbed microwave power upon magnetic resonance as well as the spin current density generated by the precessing magnetization in the vicinity of a ferromagnet/normal metal interface. In a second modeling approach, we deal with the backaction of the magnetization dynamics on the elastic wave by solving the elastic wave equation and the Landau-Lifshitz-Gilbert equation selfconsistently, obtaining analytical solutions for the acoustic wave phase shift and attenuation. We compare both modeling approaches with the complex forward transmission of a LiNbO$_3$/Ni surface acoustic wave hybrid device recorded experimentally as a function of the external magnetic field orientation and magnitude, rotating the field within three different planes and employing three different surface acoustic wave frequencies. We find quantitative agreement of the experimentally observed power absorption and surface acoustic wave phase shift with our modeling predictions using one set of parameters for all field configurations and frequencies.  
\end{abstract}




\maketitle
\section{Introduction}
Spin mechanics exploits the coupling of spin degrees of freedom with elastic properties in order to control magnetic properties and has been established in various material systems in the static and dynamic regime.\cite{S303_661,PRL97_166402,PRL99_147207,PSS2_96,NJoP10_65003,PRB78_45203,PRB79_195206,NJP11_013021,PRB81_245202,PRL105_117204,PRB82_104422,PRL106_37601}
In magnetostrictive materials, a change in the magnetization orientation results in elastic strain and vice versa, enabling strain-controlled magnetization switching.\cite{S303_661,PRB78_45203,NJP11_013021} Furthermore, an acoustic wave propagating through a ferromagnet generates elastic strains allowing to drive magnetization dynamics via the magnetoelastic interaction if the acoustic wave frequency is in resonance with the spin system at a given effective magnetic field. This magnon-phonon interaction has been investigated theoretically \cite{Tucker,JoMP5_1298,PRB7_3273,PRB7_3286,PRB9_3835} and experimentally\cite{PRL3_83,PRL7_312} in particular for bulk acoustic waves. 

Radio frequency (rf) elastic strains in ferromagnetic thin films can be generated by a surface acoustic wave (SAW) if the ferromagnetic film is deposited directly on the surface of a SAW-carrying crystal, resulting in a rigid elastic coupling. The physics of such heterostructures has been investigated experimentally.\cite{JAP47_2696,JAP53_177,JAP64_5411,JAP91_8231} Recently, it has been demonstrated that the observed changes of the complex magnetotransmission through a hybrid LiNbO$_3$/Ni SAW delay line can be identified with absorption and dispersion signals of a SAW-driven ferromagnetic resonance (FMR).\cite{PRL106_117601} 
We here use the term FMR synonymously with spin wave resonance (SWR), which is the more appropriate term if the ferromagnet is large compared with the acoustic wavelength and a spin wave mode is excited.

Since SAW-based devices are well established\cite{Datta_SAW} and can be fabricated by lithographic techniques they are attractive candidates for acoustically driven FMR in combination with ferromagnetic thin films. In acoustic FMR, the magnetization dynamics is excited by a purely internal magnetoelastic effective driving field, due to the rf strains generated by the acoustic wave in the ferromagnet.
The speed of sound is 5 orders of magnitude smaller than the speed of light, leading to a wave length of the SAW of the order of microns at GHz frequencies. This potentially allows for FMR experiments with micron-scale spatial resolution. Furthermore, spurious electromagnetic cross talk can be separated from magnetoelastic effects associated with the acoustic wave by time-domain techniques.\cite{APL72_2400,PRL106_117601} 

The latter feature is of topical interest in the context of spin pumping, i.e., the generation of a spin current by a precessing magnetization at a normal metal/ferromagnet interface that can be detected by the inverse spin Hall effect in the normal metal, e.g., a Pt layer.\cite{PRL88_117601,RMP77_1375,PRB82_214403,PRL104_46601,PRL107_46601} Since any ac electric fields which may be present at the position of the ferromagnet in a conventional FMR cavity can hamper the interpretation of the spin-pumping signal,\cite{PRB82_214403,PRL104_46601,PRL107_46601} it is important to separate electric and magnetic fields, which is naturally achieved by magnetoelastic driving as demonstrated recently.\cite{NatureMat10_737,PRL108_176601} To date, however, a thorough theoretical modeling of SAW-based FMR in combination with an extensive experimental study is still missing.

In this work, we present a theoretical framework for SAW-FMR based experiments starting from the Landau-Lifshitz-Gilbert (LLG) equation and the elastic wave equation, providing expressions that can be used to analyze SAW FMR and acoustic spin pumping experiments and are applicable to various types of SAWs (Sec.~\ref{sec:theory}). In a first approach, referred to as the ``effective field approach'', we show that a magnetoelastic driving field can be formally treated equivalently to a conventional, external ``tickle'' field (Sec.~\ref{sec:LLG}). An expression for the magnetization dynamics upon magnetoelastic driving is found, with which it is straightforward to account for the particular type of SAW employed by including the relevant strain tensor components in the magnetoelastic contribution to the free enthalpy density. Based on this expression, the power absorbed in FMR is calculated (Sec.~\ref{subsubsec:absPower}), as well as the spin current generated by the magnetization precession at a ferromagnet/normal metal interface (Sec.~\ref{subsubsec:spinCurrent}). In a second approach, referred to as the ``backaction approach'', we solve the LLG equation and the elastic wave equation simultaneously to obtain a intuitive physical picture of the backaction of the FMR on the elastic wave (Sec.~\ref{sec:backaction}). With the obtained analytical solutions, both phase shift and attenuation of the elastic wave can be calculated.
In addition, we present an extensive experimental study of SAW-FMR where the external magnetic field is rotated within three different planes employing three different SAW frequencies. We compare the experimental results with simulations based on the ``effective field approach'' and the ``backaction approach''  to discuss the advantage of each particular method (Sec.~\ref{sec:exp}). We demonstrate that the experimental data can be modeled using a single set of simulation parameters for all measurement configurations and both simulation methods. Finally, we  summarize our results and give an outlook on to further experimental and theoretical investigations (Sec.~\ref{sec:summary}).

\section{Theoretical Considerations}\label{sec:theory}
\begin{figure}[h]
\includegraphics[]{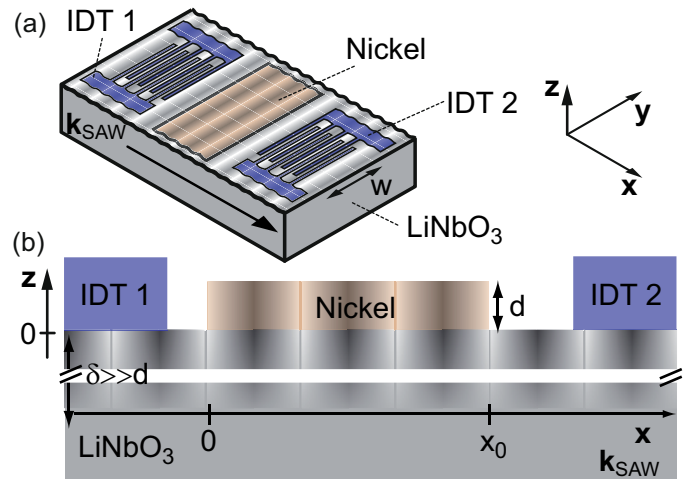}
\caption{(a) Schematic illustration of the SAW-delayline/ferromagnet hybrid device used in the experiment. The aluminum interdigital transducers (IDTs) launch and detect a SAW, which traverses the delay line, periodically straining the ferromagnetic thin film. (b) The gray shading illustrates the SAW penetrating the piezoelectric by a length $\delta$, which is of the order of the acoustic wavelength. The thickness $d$ of the ferromagnet is much smaller than the wavelength, justifying the assumption that the elastic strain within the ferromagnet is homogenous across its thickness.}
\label{fig:WaveCoupledIntoNickel}
\end{figure}

\begin{figure}[h]
\includegraphics[]{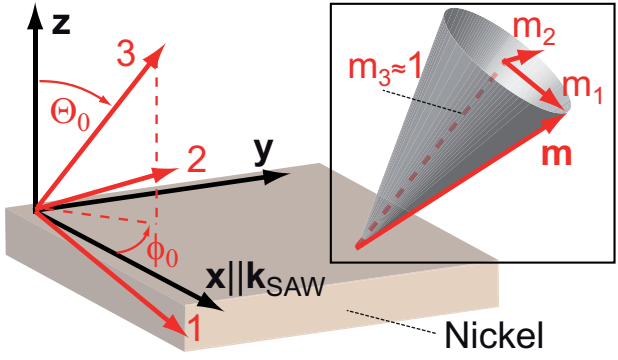}
\caption{Relation between the coordinate systems employed. The $(x,y,z)$ frame of reference is spanned by the propagation direction of the SAW, the transverse in-plane direction, and the normal of the ferromagnetic film. To solve the LLG equation, the (1,2,3) coordinate system is employed, with its 3-direction corresponding to the equilibrium magnetization orientation. The inset shows the precession cone of the magnetization, with the transverse magnetization components $m_1$ and $m_2$.}
\label{fig:CoordinateSystem}
\end{figure}

In this section, we provide the theoretical framework for the SAW-FMR experiments presented in Section \ref{sec:exp}. 
In the experiment discussed,\cite{PRL106_117601} the SAW propagates through a delay line with a ferromagnetic film deposited between two interdigital transducers (IDTs) as schematically depicted in Fig.~\ref{fig:WaveCoupledIntoNickel} and explained in more detail in Sec.~\ref{sec:exp}. We thus consider a two-layer system, consisting of a piezoelectric, single crystalline substrate and a polycrystalline, metallic ferromagnet. Rather than treating this problem numerically, we will make the following physical assumptions and simplifications to obtain analytical expressions for the acoustic wave attenuation and phase shift upon FMR as discussed in the following. The SAW penetrates the piezoelectric by a length $\delta$,\cite{Datta_SAW} of the order of the acoustic wavelength, c.f.~Fig.~\ref{fig:WaveCoupledIntoNickel}, which is typically a few microns at GHz SAW frequencies.\footnote{Note that the speed of sound in solids is of the order of 1000 m/s.} The thickness $d=50$~nm of the ferromagnetic thin film on the other hand is much smaller than the acoustic wavelength. Therefore, we assume the elastic strains within the ferromagnetic film to be homogenous within the $yz$-pane (cf.~Fig.~\ref{fig:WaveCoupledIntoNickel}), thus treating the acoustic wave traversing the ferromagnet as a bulk acoustic wave, with strain components identical to the ones at the surface ($z=0$) of the piezoelectric substrate. We focus on the acoustic wave propagation and magnetization dynamics exclusively within the ferromagnet, rather than solving the acoustic wave equation and LLG equation for a two-layer system.
We thus consider a ferromagnetic slab as depicted in Fig.~\ref{fig:CoordinateSystem} and employ a right-handed coordinate system $(x,y,z)$, where $x$ is the propagation direction of the SAW, $y$ a transverse direction, and $z$ the film normal.
In the absence of further perturbations, the direction of the magnetization $\vec{m}=\vec{M}/M$ is determined in a macro spin approach\cite{PRB72_14446} by the static free-enthalpy density (normalized to the saturation magnetization $M$) of the nickel film, given by\cite{Chikazumi}
\begin{equation}
G=-\mu_0 \vec{H}\cdot\vec{m}+B_{\rm{d}} m_z^2+B_{\rm{u}}(\vec{m}\cdot\vec{u})^2-\mu_0\vec{H}_\mathrm{ex}\cdot\vec{m},
\label{eq:G_NickelStat}
\end{equation}
where $\vec{H}$ denotes an externally applied magnetic field, $\mu_0$ is the vacuum permeability, $B_{\rm{d}}=\mu_0M/2$ represents the shape anisotropy of the thin film, $B_{\rm{u}}$ is an anisotropy parameter defining a uniaxial in-plane anisotropy along the unit vector $\vec{u}$, and $m_x$, $m_y$, and $m_z$ are the components of the unit vector $\vec{m}$; $\mu_0\vec{H}_\mathrm{ex}=D_\mathrm{s} \Delta \vec{m}$ is the exchange field with the exchange stiffness $D_\mathrm{s}$ and $\Delta=\partial_{x^2}+\partial_{y^2}+\partial_{z^2}$ is the Laplacian operator with respect to the spatial variables $x,y,z$.

In harmonic approximation, the elastic energy density $W$ is given by\cite{Tucker,Clealand_Nano} 
\begin{equation}
 W=\frac{1}{2}C_{ijkl}\varepsilon_{ij}\varepsilon_{kl},
\label{eq:stress_strain}
\end{equation}
where we make use of the Einstein summation convention. $C_{ijkl}$ denotes the tensor of the elastic constants and $\varepsilon_{ij}=(\frac{\partial u_i}{\partial x_j}+\frac{\partial u_j}{\partial x_i})/2$ with $i,j \in \{x,y,z\}$ are the strain tensor components where $u_i$ are the components of the mechanical displacement field.
The dynamic, magnetoelastic contribution to the free-enthalpy density reads\cite{Chikazumi}
\begin{eqnarray}
G^{\rm{d}}&=&b_1[\varepsilon_{xx}(x,t)m_x^2+\varepsilon_{yy}(x,t)m_y^2+\varepsilon_{zz}(x,t)m_z^2]\nonumber\\
&+& 2b_2[\varepsilon_{xy}(x,t)m_x m_y+\varepsilon_{xz}(x,t)m_x m_z\nonumber\\
&+&\varepsilon_{yz}(x,t)m_y m_z ]\label{eq:G_NickelSAW},
\end{eqnarray}
with the magnetoelastic coupling constants $b_1$ and $b_2$ and the SAW-induced dynamic strain tensor components $\varepsilon$. Equation \eqref{eq:G_NickelSAW} holds for cubic symmetry of the ferromagnetic layer; for a polycrystalline film it is further simplified by $b_1=b_2$.\cite{Chikazumi} Depending on the particular type of SAW (e.g.,~Rayleigh wave\cite{Clealand_Nano} or Love wave\cite{EL58_818}), some of the strain tensor components derived from the mechanical displacement field components\cite{Clealand_Nano} are zero as discussed in the following.
The total free enthalpy density normalized to the saturation magnetization of the film thus reads as
\begin{equation}
G^{\rm{tot}}=G+G^{\rm{d}}+W/M.
\label{eq:G_tot}
\end{equation}
Starting from this expression, we will in the following sections solve the LLG equation to obtain expressions for the magnetization dynamics and its backaction on the elastic wave in ferromagnetic resonance.

\subsection{Landau-Lifshitz-Gilbert approach\label{sec:LLG}}

The equation of motion for the magnetization direction $\vec{m}$ under the influence of an effective magnetic field $\vec{H}_\mathrm{eff}$ is the LLG equation\cite{LandauLifshitz,gilbert_phenomenological_2004}
\begin{equation}
\partial_t\vec{m}=-\gamma \vec{m}\times\mu_0 \vec{H}_{\rm{eff}}+\alpha \vec{m}\times \partial_t \vec{m},
\label{eq:LLG}
\end{equation}
where $\gamma$ and $\alpha$ are the gyromagnetic ratio and a phenomenological damping parameter, respectively. In equilibrium, the magnetization orientation is parametrized by the polar angles $\theta_0$ and $\phi_0$ for which the static free enthalpy density $G$ [Eq.~\eqref{eq:G_NickelStat}] is minimal.
We introduce a new carthesian frame of reference ($1,2,3$) in which $\vec{m}$ points along the $3$-axis and the 2-axis is in the film plane as shown in Fig.~\ref{fig:CoordinateSystem}. The transformation matrix relating the two coordinate systems depicted in Fig.~\ref{fig:CoordinateSystem} is given in Appendix~\ref{Appendix_A}.

Allowing for small deviations from the magnetization equilibrium direction $\vec{m}_0$, we write for the magnetization direction
\begin{equation}
\vec{m}=\underbrace{\left(\begin{array}{c}
     0 \\
     0\\
     1\\
     \end{array}\right)}_{\vec{m}_0}+\left(\begin{array}{c}
     m_1 \\
     m_2\\
     0\\
     \end{array}\right) +O(m_1^2,m_2^2)
\label{eq:magnetization}
\end{equation}
where $m_1$,$m_2\ll1$ in the (1,2,3) coordinate system.
The effective magnetic field is given by
\begin{equation}
\mu_0\vec{H}_{\rm{eff}}=-\nabla_{\vec{m}} G^\mathrm{tot},
\label{eq:eff_Field}
\end{equation} 
where $\nabla_{\boldsymbol{m}}=(\partial_{m1},\partial_{m2},\partial_{m3})$ is the vector differential operator
with respect to the components of $\boldsymbol{m}$. By definition of the (1,2,3) coordinate system, the static effective field is parallel to the 3-direction in the equilibrium. Following the ansatz of Baselgia {\it et al.},\cite{PRB38_2237} we expand $\nabla_{\vec{m}}G$ at the equilibrium position of $\vec{m}$, considering terms up to the first order in $m_1$ and $m_2$. For the dynamic component of the effective field, we consider only terms, which are of zeroth order in $m_1$ and $m_2$. We thus find for the total effective field
\begin{equation}
\mu_0\vec{H}_{\rm{eff}}=-\left(\begin{array}{c}
     G_{11}m_1+G_{12}m_2 \\
     G_{12}m_1+G_{22}m_2\\
     G_3\\
     \end{array}\right)
     -\left(\begin{array}{c}
     G_{1}^\mathrm{d}\\
     G_{2}^\mathrm{d}\\
     G_{3}^\mathrm{d}\\
     \end{array}\right),
\label{eq:Expansion_G}
\end{equation}
with the abbreviations $G_i=\partial_{m_i}G|_{\vec{m}=\vec{m}_0}$ and $G_{ij}=\partial_{m_i}\partial_{m_j}G|_{\vec{m}=\vec{m}_0}$ and the required explicit expressions for these derivatives given in Appendix \ref{Appendix_A}. Making a plane-wave ansatz for the transverse magnetization $m_i=m_i^0\exp[i(k x-\omega t)]$ ($i=1,2$), $\omega$ being the angular frequency, $k$ the wave number, and considering only the transverse magnetization components, we find for the LLG
\begin{equation}
    \left(\begin{array}{cc}
   G_{11}\!-\!G_3\!-\!\frac{i\omega\alpha}{\gamma} & G_{12}\!+\!\frac{i\omega}{\gamma} \\
     G_{12}\!-\!\frac{i\omega}{\gamma} & G_{22}\!-\!G_3\!-\!\frac{i\omega\alpha}{\gamma}\\
     \end{array}\right)
     \left(\begin{array}{c}
     m_1 \\
     m_2\\
     \end{array}\right)
     =
     \mu_0 
     \left(\begin{array}{c}
     h_{1} \\
     h_{2}\\
     \end{array}\right).
\label{eq:LLG_SAW_transverse}
\end{equation}
$\mu_0h_i=-G_i^\mathrm{d}=-\partial_{m_i}G^\mathrm{d}|_{\vec{m}=\vec{m}_0}$ are the components of the effective driving field 
\begin{eqnarray}
\mu_0h_1=&-&2b_1\sin\theta_0\cos\theta_0[\varepsilon_{xx}\cos^2\phi_0+\varepsilon_{yy}\sin^2\phi_0-\varepsilon_{zz}]\nonumber\\
&-&2b_2[(\varepsilon_{xz}\cos\phi_0+\varepsilon_{yz}\sin\phi_0)\cos(2\theta_0)\nonumber\\
&~&~~~~~+2\varepsilon_{xy}\sin\theta_0\cos\theta_0\sin\phi_0\cos\phi_0]\label{eq:Gd_1}\\
\mu_0h_2=&+&2b_1\sin\theta_0\sin\phi_0\cos\phi_0[\varepsilon_{xx}-\varepsilon_{yy}]\nonumber\\
&-&2b_2[\cos\theta_0(\varepsilon_{yz}\cos\phi_0-\varepsilon_{xz}\sin\phi_0)\nonumber\\
&~&~~~~~+\varepsilon_{xy}\sin\theta_0\cos(2\phi_0)].\label{eq:Gd_2}
\end{eqnarray}
In the derivation of Eq.~\eqref{eq:LLG_SAW_transverse}, we have neglected terms quadratic in $m_i$ and products of $m_i$ with the effective driving field; the latter are assumed to be of the same order as $m_i^2$.
Equation~\eqref{eq:LLG_SAW_transverse} can be solved for the transverse magnetization, resulting in
\begin{eqnarray}
 	   & &
     \left(\begin{array}{c}
     M_{1} \\
     M_{2}\\
     \end{array}\right)
     \!=
     \bar{\chi} \left(\begin{array}{c}
     h_1 \\
     h_2\\
     \end{array}\right)\label{eq:LLG_Solution} \\\nonumber
     &=&
     \!\frac{\mu_0M}{D}
    \left(\begin{array}{cc}
    G_{22}\!-\!G_3\!-\!\frac{i\omega\alpha}{\gamma}& -G_{12}\!-\!\frac{i\omega}{\gamma} \\
     -G_{12}\!+\!\frac{i\omega}{\gamma} & G_{11}\!-\!G_3\!-\!\frac{i\omega\alpha}{\gamma}\\
     \end{array}\right)
     \left(\begin{array}{c}
     h_1 \\
     h_2\\
     \end{array}\right),
\end{eqnarray}
with 
\begin{eqnarray}
D&=&(G_{11}-G_3- \frac{i\omega\alpha}{\gamma})(G_{22}-G_3-\frac{i\omega\alpha}{\gamma})\nonumber\\
&-&G_{12}^2-\left(\frac{\omega}{\gamma}\right)^2.
\label{eq:Determinant}
\end{eqnarray}
 
In Eq.~\eqref{eq:LLG_Solution}, $\bar{\chi}$ is the Polder susceptibility tensor describing the magnetic response of the ferromagnet to small time-varying magnetic fields perpendicularly oriented to $\vec{m}_0$. $\chi$ is a function of the derivatives of the static component of the free enthalpy density Eq.~\eqref{eq:G_NickelStat} while the driving field is determined by the dynamic contribution to the free enthalpy density Eq.~\eqref{eq:G_NickelSAW}. In a conventional FMR experiment, the time-varying magnetic ``tickle'' field is provided, e.g., by standing electromagnetic waves in a microwave cavity and is oriented perpendicularly to the external magnetic field. As long as the magnetization is parallel to the static external magnetic field, which is the case if the external magnetic field is large compared to the anisotropy fields, the driving-field is always perpendicular to the magnetization, independently of the magnetization orientation. This is in stark contrast to SAW-driven FMR experiments, where the driving field components $h_1$ and $h_2$ exhibit a pronounced $\vec{m}$-dependence, characteristic of the type of the SAW, i.e., of which strain component dominates the dynamic free-energy contribution Eq.~\eqref{eq:G_NickelSAW}.\cite{PRL106_117601} A Rayleigh wave contains the strain components $\varepsilon_{xx}$, $\varepsilon_{xz}$, and $\varepsilon_{zz}$,\cite{Clealand_Nano} while a Love wave is a horizontally polarized shear wave with the dominant strain component $\varepsilon_{xy}$.\cite{EL58_818} For the Rayleigh wave, at the surface the components $\varepsilon_{xx}$ and $\varepsilon_{xz}$ are phase shifted by 90~deg.
We will now discuss this $\vec{m}$-dependence of the driving field exemplarily for the strain components $\varepsilon_{xx}$,  $\varepsilon_{xy}$, and  $\varepsilon_{xz}$ separately, setting all other strain components equal to zero.
\begin{figure}[h]
\includegraphics[]{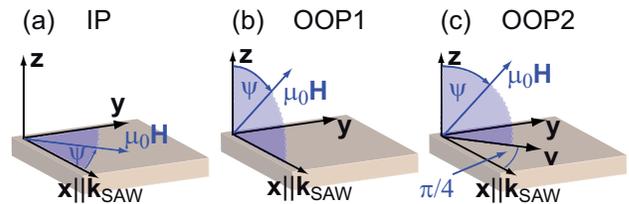}
\caption{Measurement geometries employed in the experiment. (a) In the in-plane configuration (IP) the external field is applied within the film plane. (b) and (c) In the out-of-plane configurations OOP1 and OOP2, the field is aligned in planes perpendicular to the film plane as indicated in the figure.}
\label{fig:MeasGeometries}
\end{figure}
%
%
%
To this end, we calculate the driving field $\mu_0\vec{h}$ for each of the three geometries depicted in Fig.~\ref{fig:MeasGeometries} that were used for our measurements.
 If $\vec{m}$ lies within the film plane (i.e.~$\theta_0=\pi/2$), referred to as IP configuration, we obtain the $\vec{m}$-dependent driving-field components given in the appendix \ref{Appendix_A} by Eqs.~\eqref{eq:driving_IP_eps_xx}-\eqref{eq:driving_IP_eps_xz}.
For $\vec{m}$ in the $xz$-plane (i.e.~$\phi_0=0$), referred to as OOP1 configuration, we find the expressions \eqref{eq:driving_OOP_eps_xx}-\eqref{eq:driving_OOP_eps_xz} and for the other out-of-plane configuration investigated (i.e.~$\phi_0=\pi/4$), referred to as OOP2 configuration, the driving field components are given by Eqs.~\eqref{eq:driving_OOP_2_eps_xx}-\eqref{eq:driving_OOP_2_eps_xz}.

\begin{figure}[h]
\includegraphics[]{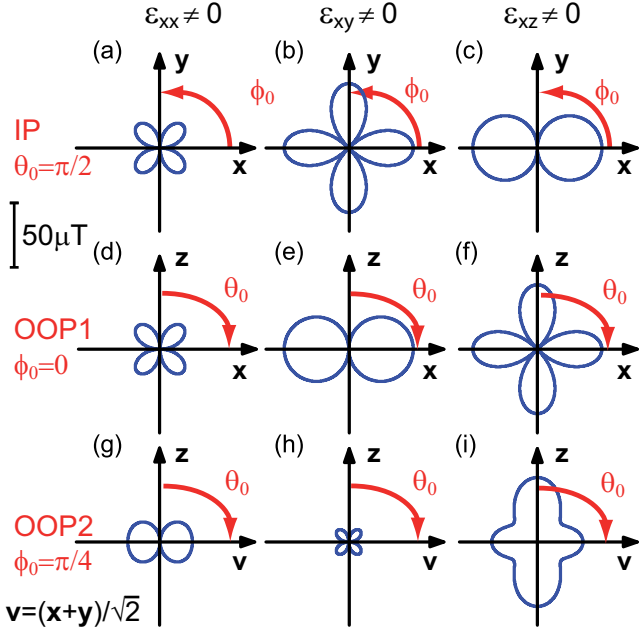}
\caption{Polar plot of the driving field's magnitude $|\mu_0 \vec{h}|$ for $\vec{m}$ in the film plane [panel (a)-(c)], out-of plane with $\phi_0=0$ [panel (d)-(f)] and with $\phi_0=\pi/4$ [panel (g)-(i)]. The distance from the origin indicates the magnitude of the driving field. The field components corresponding to the plots (a)-(i) were calculated by Eqs.~\eqref{eq:driving_IP_eps_xx}-\eqref{eq:driving_OOP_2_eps_xz} with $b_1=b_2= 25$~T and $\varepsilon_{ij} = 10^{-6}$ (see Sec.~\ref{sec:exp}); the scale for the driving field applies to all panels.}
\label{fig:DrivingPolar}
\end{figure}

The magnitude of the driving fields for the different measurement configurations and strain components is plotted in Fig.~\ref{fig:DrivingPolar} as a function of the equilibrium orientation $\vec{m}_0$ parametrized by $\phi_0$ and $\theta_0$. As Fig.~\ref{fig:DrivingPolar} demonstrates, the magnitude of the magnetoelastic driving field strongly depends on both the $\vec{m}$-orientation and the dominant strain component associated with the particular type of SAW. These characteristic fingerprints allow for a discrimination of the SAW-driven FMR from other driving mechanisms, such as electromagnetic crosstalk in free space, and for an identification of the relative amplitude of the strain components involved in the FMR excitation. As shown in Fig.~\ref{fig:DrivingPolar}, for strain amplitudes of $10^{-6}$ and magnetoelastic constants of 25~T, typically found for a SAW delayline/Ni thin film hybrid, the magnitude of $\mu_0\vec{h}$ is 50~$\mu T$.
Based on Eq.~\eqref{eq:LLG_Solution}, we will in the following subsections quantify the power, which is absorbed by the ferromagnetic film upon resonance, and the spin current density, which is generated by the precessing magnetization at a ferromagnet/normal metal interface.
\subsubsection{Absorbed Radio-Frequency Power}\label{subsubsec:absPower}
If the frequency of the SAW fulfills the ferromagnetic resonance condition, i.e., if the real part of the determinant [Eq.\eqref{eq:Determinant}] vanishes, the magnetoelastic interaction excites a resonantly enhanced magnetization precession, which in turn leads to a change in the SAW amplitude and phase. In other words, part of the SAW power is used to drive the magnetization dynamics. This is expressed in terms of a complex change of the transmitted power $\Delta P$, which we derive from Eq.~\eqref{eq:LLG_Solution} as
\begin{equation}
 \Delta P=-\frac{\omega\mu_0 }{2}\int_{V_0}\left[\left(\begin{array}{c}
     h_{1}^*, h_{2}^* \\
     \end{array}\right)\bar{\chi} \left(\begin{array}{c}
     h_{1} \\
     h_{2}\\
     \end{array}\right)\right]\mathrm{d}V,
\label{eq:CompexPower}
\end{equation}
where $V_0$ is the volume of the ferromagnetic film. 

The real and imaginary parts of $\Delta P$ are the dispersion and absorption signals associated with the FMR, respectively.\cite{PRL106_117601} Even though we have not considered an explicit backaction mechanism of the FMR on the SAW so far, Eq.~\eqref{eq:CompexPower} allows to quantitatively simulate the absorbed radio frequency power $\Im(\Delta P)$ in FMR, since by energy conservation, the power of the SAW has to be reduced by $\Im(\Delta P)$ in resonance as will be shown in Sec.~\ref{sec:exp}. The phase change of the SAW upon FMR on the other hand, is assumed to be proportional to $\Re(\Delta P)$ allowing for a qualitative discussion of the dispersion signal based on Eq.~\eqref{eq:CompexPower}. For a quantitative analysis of the phase shift of the SAW, the backaction of the magnetization dynamics on the acoustic wave is considered explicitly in Sec.~\ref{sec:backaction}.

\begin{figure}[h]
\includegraphics[]{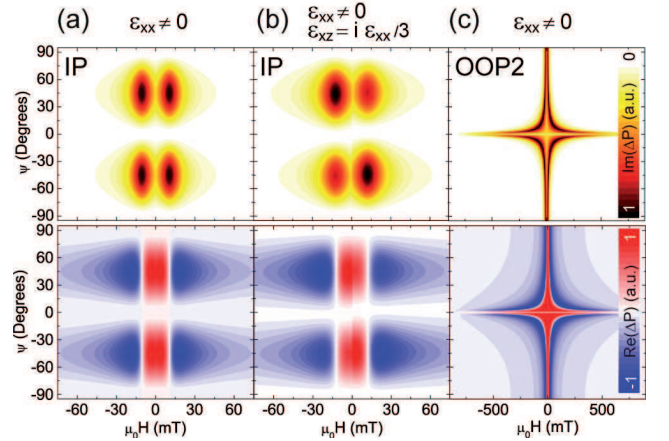}
\caption{Normalized imaginary and real parts of the complex power $\Delta P$ calculated using Eq.~\eqref{eq:CompexPower} for a pure strain along $x$ ($\varepsilon_{xx} \neq 0$) (a) and (c) and a superposition of $\varepsilon_{xx}$ and $\varepsilon_{xz}$ with $\varepsilon_{xz}$ phase shifted by $\pi/2$ (b). The magnetic field is rotated within the film plane [(a) and (b)] and perpendicular to the film plane (c); the measurement configurations and angles $\psi$ are defined in Fig.~\ref{fig:MeasGeometries}. The simulation parameters were $B_\mathrm{u}=0~$T, $B_\mathrm{d}=0.4~$T, $\omega =2\pi\times3~\mathrm{GHz}$, and $\alpha=0.1$. All other simulation parameters only affect the overall intensity of the SAW-FMR signal.} 
\label{fig:Power_IP_OOP}
\end{figure}

We will now qualitatively discuss the absorption and dispersion signals based on Eq.~\eqref{eq:CompexPower} for the IP and OOP2 configuration. In Fig.~\ref{fig:Power_IP_OOP}, $\Delta P$ is plotted as a function of the magnetic field orientation and magnitude with a specific set of parameters given in the caption. Within the film plane, $\bar{\chi}$ is independent of the field orientation for $B_\mathrm{u}=0$. Assuming a purely longitudinal strain along $x$, for the IP configuration the signature of the SAW-driven FMR directly reflects the fourfold symmetry of the magnetoelastic driving field [cf.~Fig.~\ref{fig:DrivingPolar} (a)] with the SAW-FMR signal vanishing at $\psi=0$ and $\psi=\pi/2$, c.f. Fig.~\ref{fig:Power_IP_OOP} (a). For a pure in-plane shear strain $\varepsilon_{xy}$ (not shown) the SAW-FMR signature is similar to the one in Fig.~\ref{fig:Power_IP_OOP} (a) with the difference that the SAW-FMR signal vanishes for $\psi=\pi/4$ and $\psi=3\pi/4$ [cf.~Fig.~\ref{fig:DrivingPolar} (b)]. An out-of-plane shear strain $\varepsilon_{xz}$ (not shown) results in a twofold symmetry of the SAW-FMR signal with the intensity vanishing at $\psi=\pi/2$. Depending on the type of SAW traversing the ferromagnet, various strain components with different amplitudes may superimpose, resulting in a more complex $\mu_0\vec{H}$ dependence of the FMR signal. For a Rayleigh wave traversing an elastically isotropic material there are three non-vanishing strain components at the surface of the half space, namely $\varepsilon_{xx}$, $\varepsilon_{xz}$, and $\varepsilon_{zz}$, with  a $\pi/2$ phase shift between $\varepsilon_{xx}$ and and $\varepsilon_{xz}$.\cite{Clealand_Nano} If the magnetization lies within the film plane, only the terms of the driving field containing $\varepsilon_{xx}$ and $\varepsilon_{xz}$ are relevant and superimpose with a phase factor of $\pi/2$; the resulting SAW-FMR signal is shown in Fig.~\ref{fig:Power_IP_OOP} (b), revealing an asymmetry of the SAW-FMR pattern when compared to Fig.~\ref{fig:Power_IP_OOP} (a). This asymmetry has been observed in SAW-FMR experiments,\cite{PRL106_117601,PRL108_176601} and allows to experimentally determine the relative amplitudes of the strain components associated with the SAW.
For the discussion of the out-of-plane orientations, we again focus on the longitudinal strain component and $\varepsilon_{xx}$. In the OOP1 configuration, the magnetization is oriented within the film plane along the $x$-axis for $B_\mathrm{u}=0$, due to the large demagnetization field $B_\mathrm{d}=0.4~$T (Sec.~\ref{sec:exp}), except for the case where the magnetic field points exactly along the $z$-axis. We thus expect the SAW-FMR signal to vanish for $\vec{m}$ parallel to the $x$- and $z$-axes, c.f.~Fig.~\ref{fig:DrivingPolar} (d). For the OOP2 configuration however, the driving field only vanishes for $\mu_0\vec{H}$ exactly parallel to the $z$-axis [cf.~Fig.~\ref{fig:DrivingPolar} (g)] and is nearly constant for other field orientations, since $\vec{m}$ is forced into the sample plane by the magnetic anisotropy, resulting in the SAW-FMR signal shown in Fig.~\ref{fig:Power_IP_OOP} (c).
\subsubsection{Spin-Current Generation}\label{subsubsec:spinCurrent}
At a ferromagnet/normal metal interface, the time-varying magnetization upon FMR generates an spin current density, which can be detected via the inverse spin Hall effect in the normal metal, e.g., in a  Pt layer deposited on top of the ferromagnet.\cite{PRL104_46601,PRB82_214403,PRL107_46601} It has been demonstrated that such a spin current can be driven acoustically by means of SAW-FMR.\cite{PRL108_176601} Starting from Eq.~\eqref{eq:LLG_Solution}, we derive the spin current density expected at a normal metal/ferromagnet interface in the presence of magnetoelastic driving. The spin current density reads as\cite{PRL88_117601}
\begin{equation}
j_s\vec{s}=\frac{\hbar}{4 \pi}\Re(g^{\uparrow\downarrow})\left[\vec{m} \times \frac{\partial\vec{m}}{\partial t} \right],
\label{eq:spin_current}
\end{equation}
with the spin mixing conductance $g^{\uparrow\downarrow}$.\cite{PRL104_46601} The dc component of the spin current is found by time-averaging Eq.~\eqref{eq:spin_current}\cite{PRL88_117601}
\begin{equation}
j_s^\mathrm{dc}=\frac{\hbar \omega}{8 \pi}\Re(g^{\uparrow\downarrow})\Im\left(m_1^*m_2-m_1m_2^*\right),
\label{eq:spin_current_dc}
\end{equation}
which for a circular precession $m_2=im_1$ simplifies to $j_s^\mathrm{dc}=\frac{\hbar \omega}{4 \pi}\Re(g^{\uparrow\downarrow})\sin^2\tau$,\cite{PRL88_117601} with the precession cone angle $\tau$.
Equation \eqref{eq:spin_current_dc}, together with Eq.~\eqref{eq:LLG_Solution}, allows to quantify the spin current density generated in SAW-driven FMR, where it is straight forward to account for the specific type of SAW by superimposing the relevant strain tensor components when calculating the magnetoelastic driving field Eq.~\eqref{eq:Gd_2}.

\subsection{Backaction of the ferromagnetic resonance on the acoustic wave \label{sec:backaction}}

In section \ref{sec:LLG}, we have discussed the SAW-driven FMR based on the LLG equation, which is the equation of motion for the magnetization, and used a simple energy conservation argument to quantify the absorbed SAW power upon magnetic resonance. In this section, we will consider the propagation of elastic waves through the ferromagnet and will discuss the coupling of these waves to the magnetization dynamics. 
The equation of motion of an elastic displacement with components $u_i$ is given by the elastic wave equation\cite{Tucker} 
\begin{equation}
 \rho\partial_{t^2}u_i=\partial_{x_k}\sigma_{ik}.
\label{eq:WaveEquation}
\end{equation}
The elastic displacement components are given in the ($x,y,z$) coordinate system.
In Eq.~\eqref{eq:WaveEquation}, $\rho$ denotes the mass density, and $\sigma_{ik}$ are the components of the stress tensor, which is derived from the  elastic energy density by\cite{Tucker} 
\begin{equation}
 \sigma_{ik}=\frac{\partial W}{\partial \varepsilon_{ik}}.
\label{eq:stress_strain}
\end{equation}
In the absence of magnetoelastic interactions, the elastic wave equation thus reads\cite{Tucker}
\begin{equation}
\rho\partial_{t^2}u_i=C_{ikjm}\frac{\partial^2 u_j}{\partial_{x_k}\partial_{x_m}}.
\label{eq:wave_elastic only}
\end{equation}
In the presence of magnetoelastic interaction, however, $G^\mathrm{d}M$ enters in Eq.~\eqref{eq:stress_strain} in addition to the elastic energy density; note that Eqs.~\eqref{eq:G_NickelStat} and \eqref{eq:G_NickelSAW} have been normalized to $M$. Thus, the LLG and the elastic wave equation are coupled via the magneto-elastic interaction. In the following, we will solve this coupled system of equations by making a plane-wave ansatz with the wavevector along the $x$-direction. The magnetization and the elastic displacement are thus written as $M_{1,2}=M_{1,2}^0\exp{[i(kx-\omega t)]}$ and $u_{x,y,z}=u_{x,y,z}^0\exp{[i(kx-\omega t)]}$ and the elastic mode equations linearized in $M_1$, $M_2$ read as
\begin{eqnarray}
\rho\omega^2u_x&=&c_{11}k^2u_x +2ib_1k\sin\theta_0\cos\phi_0\times \label{eq:elastic_modes}\\
&\times&\left[\sin{\phi_0}M_2-\cos\theta_0\cos\phi_0 M_1\right]\nonumber\\
\rho\omega^2u_y&=&c_{44}k^2u_y-2ib_2k\sin\theta_0 \times \nonumber\\
&\times&\left[2\sin\phi_0\cos{\phi_0}\cos{\theta_0}M_1+\cos{2\phi_0}M_2\right] \nonumber\\
\rho\omega^2u_z&=&c_{44}k^2u_z\nonumber\\
&+&2ikb_2 \left[\sin\phi_0\cos\theta_0M_2-\cos{2\theta_0}\cos\phi_0 M_1 \right]\nonumber,
\end{eqnarray}
where we have assumed cubic symmetry of the film.\cite{Tucker} For elastically isotropic media $c_{11}=(4S^2-ES)/(3S-E)$,  and $c_{44}=S$, where $E$ and $S$ are Young's modulus and the shear modulus, respectively.

Together with the linearized LLG from Eq.~\eqref{eq:LLG_SAW_transverse}, we thus obtain a system of five coupled equations with the coefficients $u_x$, $u_y$, $u_z$, $M_1$, and $M_2$. 
%
%
In order for these equations to be solved self-consistently, the determinant of the coefficients has to vanish yielding a polynomial equation in $k$.
To obtain a simple physical picture of the backaction of the magnetization dynamics on the acoustic wave, we treat the acoustic modes Eqs.~\eqref{eq:elastic_modes} separately and neglect the interaction between these modes. 

(i) First, we consider a purely longitudinal acoustic wave, i.e.~$u_y=u_z=0$. By combining Eqs.~\eqref{eq:Gd_1}, \eqref{eq:Gd_2}, \eqref{eq:LLG_Solution}, and \eqref{eq:elastic_modes}, we find
%
%
%
%
%
%
\begin{eqnarray}
\left[\omega^2\right. &-&v_\mathrm{l}^2\left( \right.1-\frac{Fb_1^2}{v_\mathrm{l}^2\mu_0\rho}\left\{\chi_{11}w_1^2 \right. \label{eq:long_backaction} \\
&+& \left. \left. \left. \chi_{22}w_2^2-(\chi_{12}+\chi_{21})w_1w_2\right\}\right)k^2\right]u_x=0\nonumber
\end{eqnarray}
where we have introduced the abbreviations $w_1=2\sin\theta_0\cos\theta_0\cos^2\phi_0$ and $w_2=2\sin\theta_0\cos\phi_0\sin\phi_0$; $v_\mathrm{l}=\sqrt{c_{11}/\rho}$ is the sound velocity of the longitudinal wave in the absence of magnetoelastic coupling. Furthermoer, we have introduced a filling factor $F<1$, which reduces the coupling of the modes, accounting for the fact that only a small fraction of the total volume traversed by the SAW is ferromagnetic, as illustrated in Fig.~\ref{fig:WaveCoupledIntoNickel}. The cross-section $A$ through which the acoustic power of the SAW is flowing is given by the penetration depth $\delta$ of the SAW times the width $w$ of the interdigital transducers. Since $\delta$ is of the order of the wavelength $\lambda$, $F=d/\delta$ will be of the order of $d/\lambda$, with the thickness $d$ of the ferromagnetic film.

(ii) Conversely, we find for a purely transverse, in-plane shear wave (i.e.,~$u_x=0$, $u_z=0$)
\begin{eqnarray}
\left[\omega^2\right. &-&v_\mathrm{t}^2\left( \right.1-\frac{Fb_2^2}{v_\mathrm{t}^2\mu_0\rho}\left\{\chi_{11}w_3^2 \right.\label{eq:ip_shear_backaction}  \\
&+& \left. \left. \left. \chi_{22}w_4^2+(\chi_{12}+\chi_{21})w_3w_4\right\}\right)k^2\right]u_y=0,\nonumber
\end{eqnarray}
with the abbreviations $w_3=2\sin{\theta_0}\cos{\theta_0}\sin{\phi_0}\cos{\phi_0}$, $w_4=\sin{\theta_0}\cos{2\phi_0}$, and the sound velocity of the transverse wave $v_\mathrm{t}=\sqrt{c_{44}/\rho}$ in the absence of magnetoelastic coupling.

(iii) For a pure out-of-plane shear wave ($u_x=0$, $u_y=0$) we find
\begin{eqnarray}
\left[\omega^2\right. &-&v_\mathrm{t}^2\left( \right.1-\frac{F2b_2^2}{v_\mathrm{t}^2\mu_0\rho}\left\{\chi_{11}w_5^2 \right.\label{eq:oop_shear_backaction}  \\
&+& \left. \left. \left. \chi_{22}w_6^2-(\chi_{12}+\chi_{21})w_5w_6\right\}\right)k^2\right]u_z=0,\nonumber
\end{eqnarray}
with $w_5=\cos\phi_0\cos{2\theta_0}$ and $w_6=\cos\theta_0\sin{\phi_0}$.

A Rayleigh type of SAW contains a longitudinal component and an out-of-plane shear component and propagates with the Rayleigh sound velocity $v_\mathrm{R}$.\cite{Clealand_Nano} To obtain a simple physical picture, we will ignore the transverse component and assume the wave to be purely longitudinal with $v_\mathrm{l}=v_\mathrm{R}$. This simplification is motivated by the observation that most characteristic features of the angle dependent SAW-FMR measurements can be simulated assuming a magnetoelastic driving field with a purely longitudinal strain. \cite{PRL106_117601}

Assuming that the changes of the mode due to magnetoelastic interactions are small and neglecting the exchange interaction [$D_\mathrm{s}=0$ in Eq.~\eqref{eq:G_3}], we find for the wave number of the perturbed longitudinal wave Eq.~\eqref{eq:long_backaction}
%
%
%
%
%
\begin{eqnarray}
k&=&k_0+\Delta k \label{eq:k_long}\\
\Delta k &=& F\frac{\omega b_1^2}{2v_\mathrm{R}^3\mu_0\rho}\{\chi_{11}w_1^2 + \chi_{22}w_2^2-(\chi_{12}+\chi_{21})w_1w_2\}),\nonumber
\end{eqnarray}
with the unperturbed wavenumber $k_0=\omega/v_\mathrm{R}$. Note that at the largest SAW frequency employed in this work (2.24~GHz) the exchange term $D_\mathrm{s}k^2$ with $D_\mathrm{s}=2.1\times 10^{-17}~\mathrm{Tm}^2$ (Ref.~\onlinecite{JAP53_2029}) results in an isotropic resonance field shift of only 0.4~mT justifying the disregard of $D_\mathrm{s}$.
%


With Eq.~\eqref{eq:k_long} we thus find for the longitudinal acoustic wave in the presence of magnetoelastic interaction%
\begin{equation}
u_x(x=x_0)=u_{x}(x=0)e^{ik_0x_0}\underbrace{\exp{( i\Delta k x_0)}}_{S_{21}^{\mathrm{norm}}},
\label{eq:Rayleigh_long_S21}
\end{equation}
where $x_0$ is the length of the ferromagnetic film. In Eq.~\eqref{eq:Rayleigh_long_S21}, we have defined a normalized complex scattering parameter $S_{21}^{\mathrm{norm}}$, which describes the resonant attenuation and phase shift of the acoustic wave with $S_{21}^{\mathrm{norm}}=1$ off resonance, c.f.~Eq.~\eqref{eq:k_long}.

The power associated with the acoustic wave can be derived from its Pointing vector and reads as\cite{Clealand_Nano}
\begin{equation}
P_\mathrm{ac}=\underbrace{\frac{1}{2}A \rho v_\mathrm{R}\omega^2 u_{x}(x=0)^2}_{P_\mathrm{ac}^0} |S_{21}^{\mathrm{norm}}|^2,
\label{eq:Rayleigh_long_Power}
\end{equation}
where $P_\mathrm{ac}^0$ is the acoustic power of the wave out of resonance.
For small deviations of $|S_{21}^{\mathrm{norm}}|^2=e^{-2\Im{(\Delta k)}x_0}$ from unity, the change of the acoustic power in ferromagnetic resonance can be shown to be identical to the imaginary part of Eq.~\eqref{eq:CompexPower} setting all strain components but $\varepsilon_{xx}=ik u_{x}$ to zero. Accordingly, the real part of Eq.~\eqref{eq:CompexPower} can be related to the phase $\arg(S_{21}^{\mathrm{norm}})=\Re{(\Delta k)}x_0$. We thus find
\begin{equation}
\Delta P =-2\Delta k x_0 P_\mathrm{ac}^0,
\label{eq:relation_complex_power_backaction}
\end{equation}
demonstrating the consistency of the two models discussed in Secs.~\ref{sec:LLG} and \ref{sec:backaction}. We stress that this relation holds for small perturbations $\Delta k$ of the wavenumber and for a purely longitudinal mode. Particularly, larger phase changes of the order of $\pi/2$ as observed in the experiment cannot be accounted for within this modeling approach because they would lead to $\Re(\Delta P) > P_\mathrm{ac}^0$.

\subsubsection{Magneto-phonon polaritons}
\begin{figure}[!htbp]
\includegraphics[]{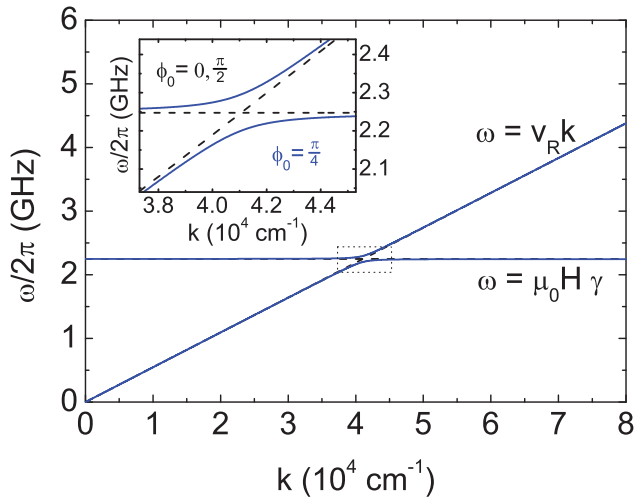}
\caption{Dispersion relation of the acoustic wave and the FMR mode, calculated by solving Eq.~\eqref{eq:anticrossing} with the parameters given in the text. The dashed lines correspond to the FMR mode and the acoustic wave for $\phi_0=0$ and $\phi_0=\pi/2$, where the coupling term in Eq.~\eqref{eq:anticrossing} vanishes. The solid lines represent the modes for $\phi_0=\pi/4$ where the coupling of the modes is strongest, manifesting itself in the avoided level crossing.}
\label{fig:CoupledModes}
\end{figure}

To illustrate the coupling of the elastic wave and the magnetization dynamics, we consider Eq.~\eqref{eq:long_backaction} with the magnetization oriented within the plane, i.e., $\theta_0=\pi/2$; for simplicity, we assume $B_\mathrm{d}=B_\mathrm{u}=0$, and $\alpha=0$. Equation~\eqref{eq:long_backaction} simplifies to
\begin{equation}
\omega^2-v_\mathrm{R}^2k^2\left(1-\frac{4b_1^2\sin^2\phi_0\cos^2\phi_0\mu_0H \mu_0M}{v_\mathrm{R}^2\mu_0\rho\left((\mu_0H)^2-\left(\frac{\omega}{\gamma}\right)^2\right)}\right)=0
\label{eq:anticrossing}
\end{equation}
%
%
%
%
which can readily be solved for $\omega$.
The magnetic field-dependent part of Eq.~\eqref{eq:anticrossing} is proportional to the square of the driving field for $\vec{m}$ within the plane and only $\varepsilon_{xx}\neq 0$, cf.~Eq.~\eqref{eq:driving_IP_eps_xx}. Considering Fig.~\ref{fig:DrivingPolar} (a), we expect the strongest interaction of the FMR mode with the elastic wave for $\phi_0=\pi/4$ and no interaction for $\phi_0=0$ and $\phi_0=\pi/2$. We assume a magnetic field of $\mu_0H=73.5$~mT, corresponding to a resonance frequency of 2.24~GHz, i.e., the highest SAW frequency employed in this work, and use typical parameters for nickel with the references given below; $v_\mathrm{R}=3440~$m/s, $d=50$~nm, $b_1=23$~T, $\rho=8900$~kg/m$^3$, $\gamma=2.185\mu_\mathrm{B}/\hbar$ with Bohr's magneton $\mu_\mathrm{B}$ and the reduced Planck constant $\hbar$. The wavelength of the acoustic wave at the expected FMR position is $\lambda_\mathrm{res}=2\pi v_\mathrm{R}/(\mu_0 H \gamma)= v_\mathrm{R}/(2.24~\mathrm{GHz})$ and we therefore assume a filling factor of $F=d/\lambda_\mathrm{res}$. We solve Eq.~\eqref{eq:anticrossing} for $\omega$ and plot in Fig.~\ref{fig:CoupledModes} the obtained result against $k$ for $\phi_0=0$, $\phi_0=\pi/4$, and $\phi_0=\pi/2$. As long as the acoustic wave frequency is off-resonant with the FMR frequency at the given field, we observe a linear dispersion $\omega = v_\mathrm{R}k$ of the acoustic wave and a $k$-independent mode at $\omega=\mu_0H\gamma$. Because there is no coupling of these modes for  $\phi_0=0$ and $\phi_0=\pi/2$, we observe a mode crossing. For $\phi_0=\pi/4$, however, the coupling of the modes is strongest and we observe a mode hybridization, manifesting itself in an avoided level crossing, as shown in Fig.~\ref{fig:CoupledModes}; these hybridized modes are referred to as magneto-phonon polaritons.\cite{JAP108_013520}  As it can be seen in Fig.~\ref{fig:CoupledModes}, the mode we are referring to as FMR mode is $k$-independent and exhibits the avoided level crossing with the acoustic wave mode at $k\neq0$, which is why it is more correct to refer to this mode as SWR mode.

Similar to the modeling of a Rayleigh wave by a longitudinal bulk mode discussed above, the attenuation and phase shift of other SAW modes in FMR can be modeled taking into account the transverse bulk modes. A SAW-FMR experiment with a Love wave, e.g., can be modeled by an in-plane shear wave Eq.~\eqref{eq:ip_shear_backaction}.

\section{Experiment and Discussion\label{sec:exp}}

Having established the theoretical framework of SAW-FMR, we turn to the experiment. We investigate the hybrid SAW delay line device schematically depicted in Fig.~\ref{fig:WaveCoupledIntoNickel}, fabricated from a $y$-cut $z$-propagation LiNbO$_3$ substrate. The 70~nm-thick aluminum transducers with an interdigital spacing of 5~$\mu$m and a metallization ratio of 50\% were fabricated using optical lithography and e-beam evaporation. This geometry results in a fundamental frequency of $172$~MHz, corresponding to a sound velocity of $v_\mathrm{R}=3440$~m/s,\cite{PRL106_117601} and yields a bandpass at odd harmonic frequencies; for the SAW-FMR experiments, frequencies of 0.86~GHz, 1.55~GHz, and 2.24~GHz were employed. A polycrystalline Ni film with dimensions $d=50$~nm, $w=400~\mu$m, and $x_0=570~\mu$m was deposited between the transducers by e-beam evaporation.
The complex forward transmission $S_{21}$ of the delay line, defined as the voltage ratio of the electromagnetic wave detected and applied at IDT 2 and 1, respectively, was measured using vector network analysis with an input power of $P=0.1$~mW. The SAW-transmission was isolated from spurious signals such as electromagnetic cross-talk and multiple transit signals by Fourier transformation and time gating.\cite{APL72_2400,PRL106_117601} The delay line was mounted between the poles of a rotatable electromagnet and measurements were carried out in the three different measurement configurations shown in Fig.~\ref{fig:MeasGeometries}. All experiments were performed at room temperature. For each harmonic frequency, the power of the SAW was determined by measuring the scattering parameter $S_{11}$, defined as the voltage ratio of the reflected and applied electromagnetic wave at the transducer. The fraction of the applied electromagnetic power coupled into the SAW is given by $P_\mathrm{SAW}=\Delta|S_{11}|^2 P_0/2 $, where $\Delta|S_{11}|^2$ is the amplitude of the dip occurring at the corresponding harmonic center frequency of the transducer and $P_0$ is the input power applied to the transducer; the factor 1/2 accounts for the transducer bidirectionality. The change of phase and amplitude of the SAW upon ferromagnetic resonance was determined by normalizing the scattering parameter $S_{21}^{\mathrm{norm}}=S_{21}(\mu_0H)/S_{21}(\mu_0H_\mathrm{off})$, where the off-resonance field $\mu_0 H_\mathrm{off}=150$~mT for the in-plane measurements and $\mu_0 H_\mathrm{off}=1.2$~T for the out-of-plane configurations. The resonantly absorbed power was calculated from the normalized measured scattering parameter by $P_\mathrm{abs}=(1-|S_{21}^{\mathrm{norm}}|^2)P_\mathrm{SAW}$.

We will now compare the experimental data obtained for the absorbed power $P_\mathrm{abs}$ with simulations based on the imaginary part of Eq.~\eqref{eq:CompexPower}, which has the advantage that different strain components can be easily incorporated in the simulation. Furthermore, we will compare the measured normalized scattering parameter $S_{21}^{\mathrm{norm}}$ with the one simulated with Eq.~\eqref{eq:Rayleigh_long_S21}, i.e., describing attenuation and phase shift of a purely longitudinal mode.
\begin{figure*}[!htbp]
\includegraphics[]{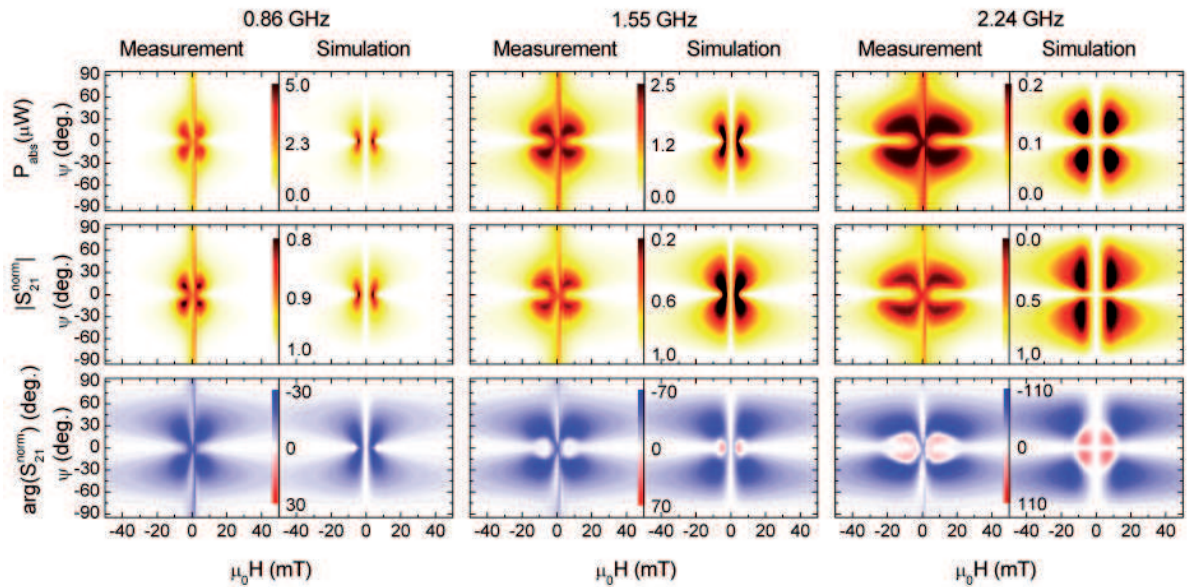}
\caption{Experiment and simulation of the angle-dependent SAW-FMR in the IP configuration. The definition of the angle $\psi$ is given in Fig.~\ref{fig:MeasGeometries} (a). In the top row the power absorbed upon FMR is shown for all employed frequencies. For the simulation the imaginary part of Eq.~\eqref{eq:CompexPower} was used with all parameters given in the text and in Tab.~\ref{tab:Parameters}. In the second and third row, the measured magnitude and phase of the normalized scattering parameter $S_{21}^{\mathrm{norm}}$ are shown together with the corresponding simulation, using Eq.~\eqref{eq:Rayleigh_long_S21}. }
\label{fig:ExpSimu_IP}
\end{figure*}
%
%
%
%
%
%
%

We start with the in-plane configuration where $\mu_0\vec{H}$ was rotated within the film plane. Here, $\psi$ denotes the angle between the SAW propagation direction and $\mu_0\vec{H}$ as defined in Fig.~\ref{fig:MeasGeometries}. For each employed frequency, the measured and simulated data are shown in a false color plot with the identical scale for simulation and experiment. The top row shows the absorbed power, simulated with Eq.~\eqref{eq:CompexPower}, i.e., the ``effective field approach''. The best agreement between experiment and simulation for all frequencies was found for the parameters $\alpha=0.1$, $B_\mathrm{d}=400~$mT, $B_\mathrm{u}=2.5~$mT, and $D_\mathrm{s}=0$; the exchange stiffness was neglected for reasons discussed above. The value for $\alpha$ is about a factor of two larger than literature values for polycrystalline nickel thin films.\cite{JPD41_164016} We conjecture that other, non-Gilbert type, line broadening mechanisms play a role. These mechanisms could include two-magnon processes,\cite{PRB60_7395} non-uniform excitation due to the small acoustic wave length,\cite{JAP95_5646}, non-resolved standing spin-wave modes,\cite{JAP48_382,Hoekstra,APL82_730,PRB79_45205} or excitation of other spin-wave modes and their damping. The clarification of this issue would require a systematic, frequency-dependent study of the SAW-FMR linewidth, particularly at higher SAW frequencies where frequency-independent contributions to the linewidth are less dominant and where standing spin-wave modes can be spectrally resolved. However, this would require the fabrication of higher frequency SAW devices by electron-beam lithography, which is beyond the scope of this work.

The magnitude of the strain components derived from the simulations depends on the frequency with the respective values given in Tab.~\ref{tab:Parameters}; for higher frequencies, a larger fraction of the electromagnetic power applied to the transducer is emitted into the free space and therefore less electromagnetic power is transfered into the SAW. As further simulation parameters, the following literature values were used:  $M=370$~kA/m (Ref.~\onlinecite{NJP11_013021}), $b_1=b_2=-3\lambda_\mathrm{s}c_{44}=23$~T (Ref.~\onlinecite{Chikazumi}), with the isotropic magnetostriction constant $\lambda_\mathrm{s}=-38\times 10^{-6}$ (Ref.~\onlinecite{JAP53_2661}) and the elastic shear module $c_{44}=S=74$~GPa (Ref.~\onlinecite{PR77_566}), $\rho=8900$kg/m$^3$ (Ref.~\onlinecite{Mills}), and $\gamma=2.185\mu_\mathrm{B}/\hbar$ (Ref.~\onlinecite{JAP32_330}).
As shown by the graphs, the characteristic angle-dependence as well as the absolute value of the absorbed power of the experimental data is reproduced by the simulation with one set of parameters. 

In the second and third row of Fig.~\ref{fig:ExpSimu_IP}, the magnitude and phase of the measured normalized $S_{21}^{\mathrm{norm}}$ parameter are shown together with the simulated $S_{21}^{\mathrm{norm}}$ using Eq.~\eqref{eq:Rayleigh_long_S21}, i.e., the ``backaction approach''. For this simulation, the same parameters as above were used. The strain components do not enter in Eq.~\eqref{eq:Rayleigh_long_S21}, but $F$ is used as a free parameter, which is of the order of $d/\lambda$ and therefore depends on the frequency; the frequency-dependent parameters used for the simulations are also summarized in Tab.~\ref{tab:Parameters}. In the simulation, $F$ was chosen such that the simulated phase $\arg(S_{21}^{\mathrm{norm}})$ quantitatively agrees with the experiment. In the SAW-FMR dispersion, the characteristic angle-dependence is reproduced in the simulation. With this set of parameters, the simulated magnitude $|S_{21}^{\mathrm{norm}}|$ is also in reasonably good agreement with the experiment, given the simplicity of the model.\footnote{We emphasize that the two approaches are equivalent only for small perturbations of the modes which is not the case here.} Note that the nearly field-orientation independent change in the magnetotransmission observed in the absorption and dispersion data at low positve fields stems from hysteretic magnetization switching,\cite{JAP53_177,PRL106_117601} and can be included in the modeling by searching for the local energy minimum of the free enthalpy density Eq.~\eqref{eq:G_NickelStat} with respect to the magnetization direction rather than the global minimum. However, we here focus on the non-hysteretic SAW-FMR signature and thus disregard the hysteretic switching in the simulation.

Particularly at 2.24~GHz, the simulated maximum value for $|S_{21}^{\mathrm{norm}}|$ overestimates the acoustic wave attenuation observed in the experiment. Most likely, the agreement could be improved by modeling the Rayleigh wave more realistically by considering the attenuation and phase shift of a coupled longitudinal and transverse wave.  
Since $|S_{21}^{\mathrm{norm}}|$ substantially deviates from unity in the experiment, the approximation discussed in Sec.~\ref{sec:backaction} when showing that the change of acoustic power for the longitudinal mode upon FMR is identical to the imaginary part of Eq.~\eqref{eq:CompexPower} is not justified. Nevertheless, the modeling of the absorbed power with $\Im(\Delta P)$ can reproduce the experimentally observed lineshape of the absorption data better than the modeling with Eq.~\eqref{eq:Rayleigh_long_S21}, since all relevant strain components can easily be accounted for. As already mentioned in Sec.~\ref{sec:backaction}, the experimentally observed phase shift of more than 90~deg., however, can not be quantitatively modeled with Eq.~\eqref{eq:CompexPower}. We finally note that, within both approaches, for the magnetic field orientations close to the SAW propagation direction, i.e., for $\psi$ close to 0~deg.~in Fig~\ref{fig:ExpSimu_IP}, the agreement between experiment and simulation is less good than for other orientations. This could be a consequence of the simplifications involved in the modeling and might be improved by considering non-linear driving.\cite{JMMM272_1009,TP47_38}

\begin{figure*}[!htbp]
\includegraphics[]{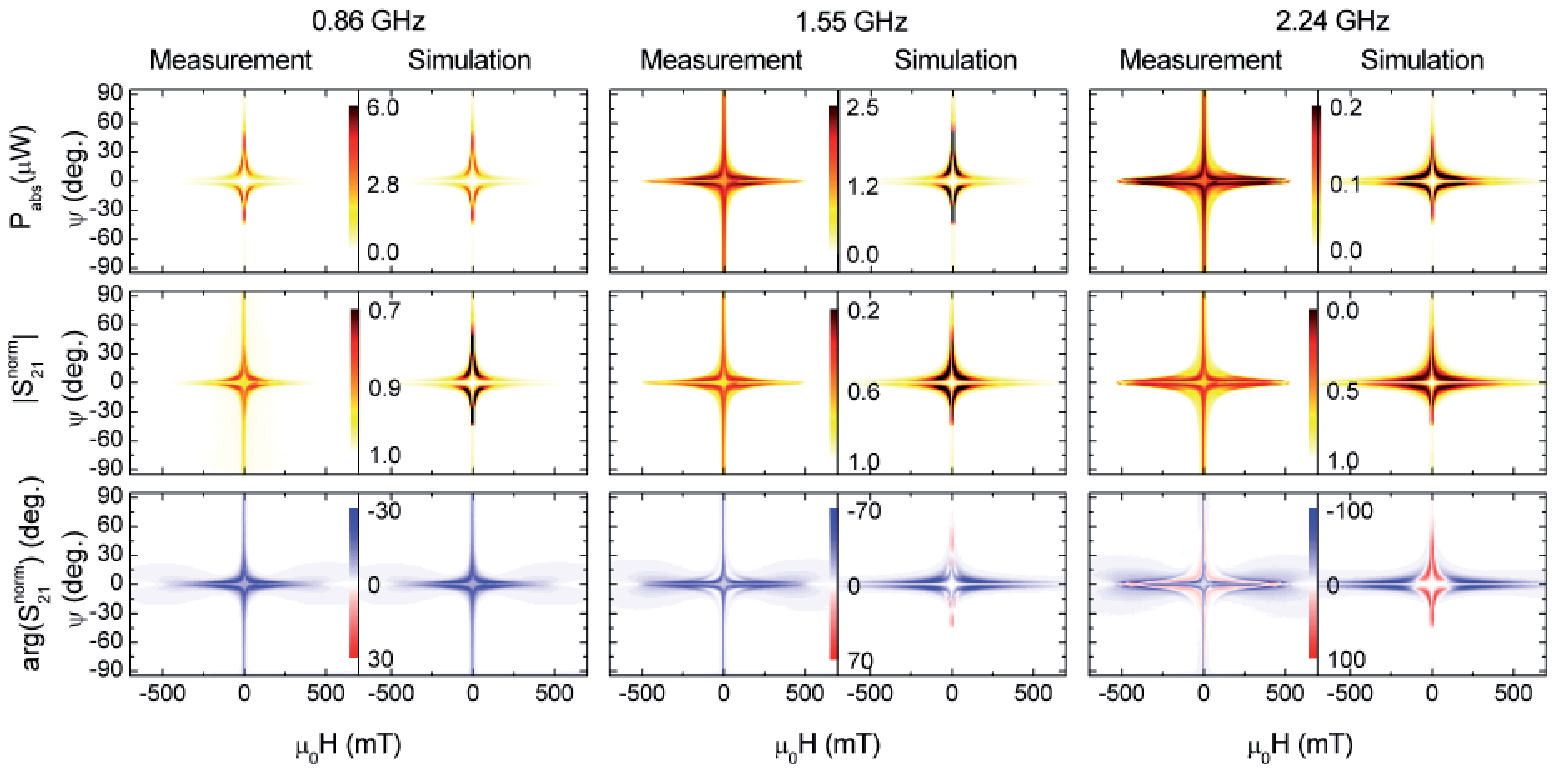}
\caption{Experiment and simulation of the angle-dependent SAW-FMR in the OOP1 configuration. The definition of the angle $\psi$ is given in Fig.~\ref{fig:MeasGeometries} (b). In the top row the power absorbed upon FMR is shown for all employed frequencies. For the simulation the imaginary part of Eq.~\eqref{eq:CompexPower} was used with the same parameters as in the IP configuration. In the second and third row, the measured magnitude and phase of the normalized scattering parameter $S_{21}^{\mathrm{norm}}$ are shown together with the corresponding simulation, using Eq.~\eqref{eq:Rayleigh_long_S21}.}
\label{fig:ExpSimu_OOP1}
\end{figure*}

\begin{figure*}[!htbp]
\includegraphics[]{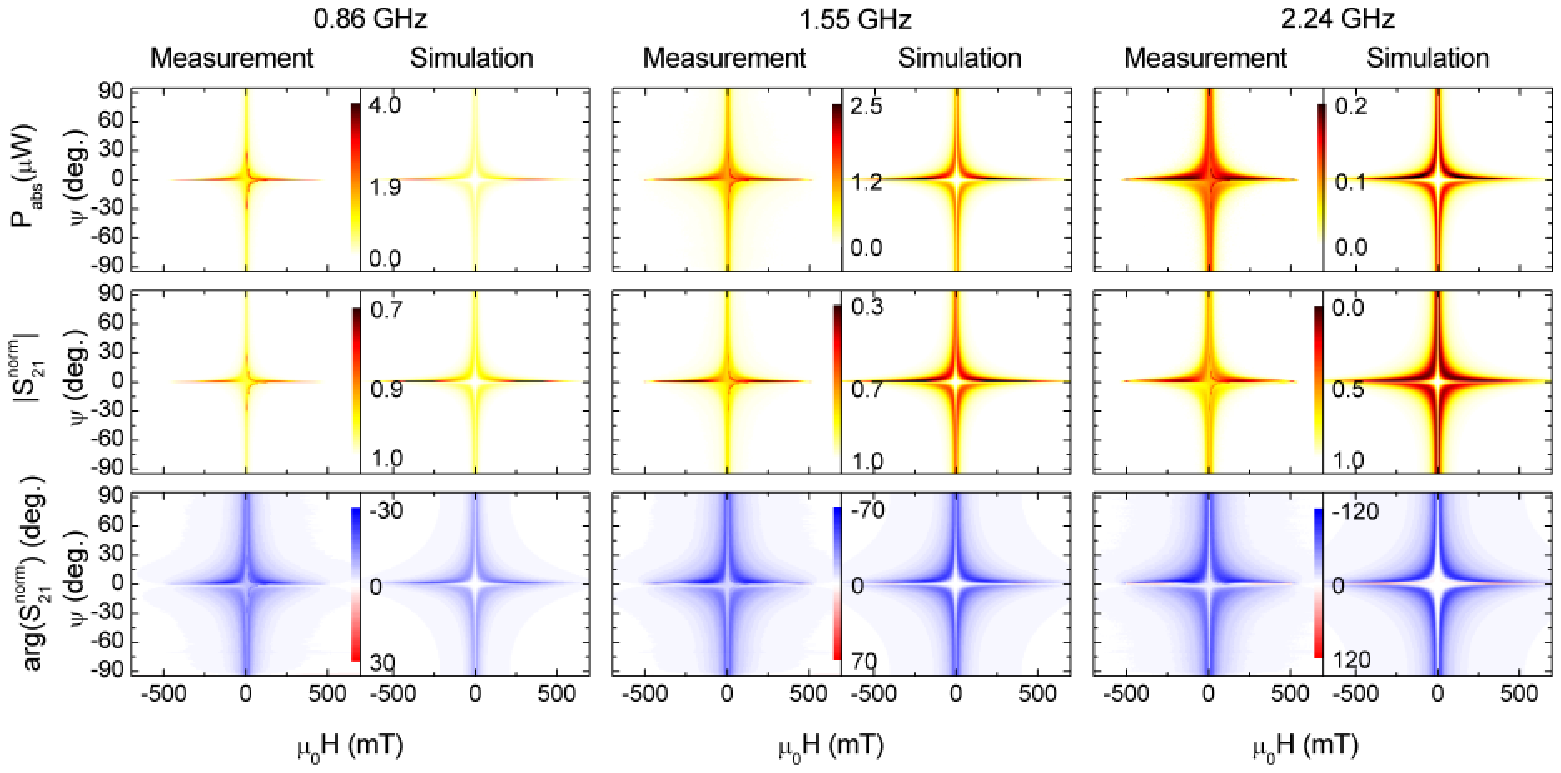}
\caption{Experiment and simulation of the angle-dependent SAW-FMR in the OOP2 configuration. The definition of the angle $\psi$ is given in Fig.~\ref{fig:MeasGeometries} (c). In the top row the power absorbed upon FMR is shown for all employed frequencies. For the simulation the imaginary part of Eq.~\eqref{eq:CompexPower} was used with the same parameters as in the IP configuration. In the second and third row, the measured magnitude and phase of the normalized scattering parameter $S_{21}^{\mathrm{norm}}$ are shown together with the corresponding simulation, using Eq.~\eqref{eq:Rayleigh_long_S21}.}
\label{fig:ExpSimu_OOP2}
\end{figure*}

\begin{table}[h]
\flushleft
\centering
\caption{Frequency-dependent parameters used for the simulations in Figs.~\ref{fig:ExpSimu_IP}-\ref{fig:ExpSimu_OOP2}.}
\begin{ruledtabular}
\begin{tabular}{l c c c }
~& $0.86~$GHz & $1.55~$GHz & $2.24~$GHz \\\hline
$\varepsilon_{xx}~(10^{-6})$ &1.8&1.15&0.36\\
$\varepsilon_{xz}~(10^{-6})$ &0.09i&0.0575i&0.018i\\
$\varepsilon_{zz}~(10^{-6})$ &0.18&0.115&0.036\\
$F~(d/\lambda)$ &0.26&0.35&0.38\\
$\lambda$~($\mu$m)&20/5&20/9&20/13\
\label{tab:Parameters}
\end{tabular}
\end{ruledtabular}
\end{table}
Figures \ref{fig:ExpSimu_OOP1} and \ref{fig:ExpSimu_OOP2} show the experimental and simulated data for the OOP1 and OOP2 field orientations, with the angle $\psi$ defined in Fig.~\ref{fig:MeasGeometries} and the panels organized in the same way as in Fig.~\ref{fig:ExpSimu_IP}. For the simulations the identical parameters as for the IP configuration were used, resulting again in a good agreement between experiment and simulation. For both configurations a slight asymmetry along the $\psi$-axis is observed. It stems from a misalignment of the sample with respect to the magnetic field of less than one degree and is reproduced in the simulation. 
We finally note that the overall agreement between experiment and theory for the absorption data is slightly better in the ``effective field approach'' than in the ``backaction approach'', particularly the quantitative value of the absorbed power can be reproduced very well in this approach, while the ``backaction approach'' is better suitable for describing the phase.

\section{Summary and Outlook \label{sec:summary}}
In summary, we have provided a theoretical framework for SAW-based FMR experiments based on two different approaches. In the ``effective field approach'', we calculated the magnetization dynamics in the presence of a magnetoelastic driving field. Based on the obtained analytical expression, the rf power absorption in the ferromagnetic film can be calculated under magnetic resonance conditions. The advantage of this method is that arbitrary types of SAWs or bulk waves which may drive the ferromagnetic resonance can be modeled in a rather simple fashion. The modeling is achieved by superimposing the corresponding strain components in the magnetoelastic terms of the free enthalpy density, where necessary with an additional phase factor. Further, we have derived an expression for the spin-current generation upon FMR, which applies to SAW-based acoustic spin-pumping experiments in ferromagnet/normal metal hybrids, again with the capability of modeling arbitrary acoustic wave modes.

In the ``backaction approach'', we have taken into account the backaction of the magnetization dynamics on the acoustic wave by solving the LLG equation and the bulk elastic wave equation, neglecting the interaction between the longitudinal and transverse bulk waves. Despite the approximations involved, the analytical results allow a quantitative analysis of the phase shift and attenuation of the SAW upon FMR and we showed that for small perturbations of the longitudinal elastic wave the two approaches are consistent. The particular advantage of this approach is that the SAW phase shift in FMR can be modeled quantitatively.

Furthermore, we have performed systematic SAW-FMR experiments using a Ni/LiNbO$_3$ hybrid SAW delayline and measuring the complex scattering parameter of the device $S_{21}^{\mathrm{norm}}$ as a function of the magnetic field orientation and magnitude; the field was rotated in the in-plane configuration and in two different out-of-plane configurations for three different SAW frequencies each.
We have shown that the absorbed power can be quantitatively described with the LLG approach with one parameter set for all measurement configurations and frequencies. Moreover, we have shown that the phase shift of the SAW upon FMR can be quantitatively modeled by the backaction model, considering a longitudinal bulk wave traversing through the hybrid.

This work thus lays theoretical foundations for SAW-based spin mechanics experiments such as SAW-FMR \cite{PRL106_117601} driven acoustic spin pumping \cite{PRL108_176601} and is applicable to various (surface) acoustic wave modes. 
Further theoretical work could be directed toward describing the SAW propagation through the coupled two-layer system consisting of the piezoelectric substrate and the ferromagnetic layer numerically, i.e., without the approximations discussed in Sec.~\ref{sec:theory}, to obtain an even more accurate description of the experiment and quantitative modeling of the SAW attenuation and phase shift. Particularly, it would be desirable to improve the agreement between theory and experiment for field orientations close to the propagation direction of the SAW. A particularly appealing experiment would be to spectroscopically resolve the avoided level crossing of the SAW and FMR mode. To this end, a ferromagnet with a FMR linewidth of the order of 10~MHz with a similar magnetoelastic coupling as Ni or SAW devices with higher frequencies and thus larger filling factors would be required. Therefore, it could be beneficial to intentionally use standing spin-wave resonances, e.g.~by employing magnonic crystals, and couple these to a SAW resonator. Alternatively, the coupling of paramagnetic centers to elastic waves,\cite{PRL107_235502} e.g. via a crystal field,\cite{Tucker} could be exploited to observe a strong coupling of the spin ensemble to a SAW resonator, in analogy to the strong coupling of spin ensembles to photon cavities.\cite{PRL105_140501,PRL105_140502,PRL105_140503}

\begin{acknowledgments}
This work was supported by the Deutsche Forschungsgemeinschaft via SFB631 project C3, Cluster of Excellence Nanosystems Initiative Munich (NIM), and SPP 1538 ``Spin Caloritronic Transport'' project GO 944/4-1. \\
We greatfully acknowledge discussions with Akashdeep Kamra.
\end{acknowledgments}

\appendix
\section{Coordinate transformation and free enthalpy derivatives\label{Appendix_A}}
The transformation between the ($x,y,z$) coordinate system, defined by the propagation direction of the SAW $x$ and the surface normal $z$, and the equilibrium system (1,2,3) is given by
\begin{equation}
     \left(\begin{array}{c}
     m_x \\
     m_y\\
     m_z\\
     \end{array}\right)=U \left(\begin{array}{c}
     m_1 \\
     m_2\\
     m_3\\
     \end{array}\right),
    \label{eq:Transformation}
\end{equation}
with
\begin{equation}
     U=\left(\begin{array}{ccc}
     \cos\theta_0\cos\phi_0 & -\sin\phi_0 & \sin\theta_0\cos\phi_0 \\
     \cos\theta_0\sin\phi_0 & \cos\phi_0 &\sin\theta_0\sin\phi_0\\
     -\sin\theta_0 & 0 & \cos\theta_0 \\
     \end{array}\right).
    \label{eq:TransformationMatrix}
\end{equation}

The driving fields derived from the derivatives of the static free-enthalpy density Eq.~\eqref{eq:G_NickelStat} with respect to the magnetization components are given by
\begin{eqnarray}
G_3&=&\partial_{m_3}G|_{\vec{m}=\vec{m_0}}\\\nonumber
&=&-\mu_0 H_3+2B_\mathrm{d}\cos^2\theta_0+2B_\mathrm{u}u_3^2+D_\mathrm{s}k^2,\\
\label{eq:G_3}
G_{21}&=&G_{12}=\partial_{m_1}\partial_{m_2}G|_{\vec{m}=\vec{m_0}}=2B_uu_2u_1, \label{eq:G_12}\\
G_{11}&=&\partial^2_{m_1}G|_{\vec{m}=\vec{m_0}}=2B_\mathrm{d}\sin^2\theta_0+2B_uu_1^2, \label{eq:G_11}\\
G_{22}&=&\partial^2_{m_2}G|_{\vec{m}=\vec{m_0}}=2B_uu_2^2.\label{eq:G_22}
\end{eqnarray}
In the following, the derivatives of the dynamic free enthalpy density used for the plots in Fig.~\ref{fig:DrivingPolar} are stated explicitly for the strain components $\varepsilon_{xx}$,  $\varepsilon_{xy}$, and  $\varepsilon_{xz}$ separately, setting all other strain components equal to zero. If $\vec{m}$ lies within the film plane (i.e.~$\theta_0=\pi/2$), referred to as IP configuration, we obtain the $\vec{m}$-dependent driving-field components
\begin{equation}
\mu_0\left(\begin{array}{c}
     h_{1} \\
     h_{2}\\
     \end{array}\right)
     =
     \left(\begin{array}{c}
     0 \\
     2b_1\varepsilon_{xx}\sin\phi_0\cos\phi_0\\
     \end{array}\right)\label{eq:driving_IP_eps_xx},
\end{equation}
\begin{equation}
\mu_0\left(\begin{array}{c}
     h_{1} \\
     h_{2}\\
     \end{array}\right)
     =
     \left(\begin{array}{c}
     0 \\
     -2b_2\varepsilon_{xy}\cos(2\phi_0)\\
     \end{array}\right)\label{eq:driving_IP_eps_xy},
\end{equation}
and
\begin{equation}
\mu_0\left(\begin{array}{c}
     h_{1} \\
     h_{2}\\
     \end{array}\right)
     =
     \left(\begin{array}{c}
     2b_2\varepsilon_{xz}\cos\phi_0 \\
     0\\
     \end{array}\right).\label{eq:driving_IP_eps_xz}
\end{equation}

Conversely, if $\vec{m}$ is in the $xz$-plane (i.e.~$\phi_0=0$), referred to as OOP1 configuration, we find
\begin{equation}
\mu_0\left(\begin{array}{c}
     h_{1} \\
     h_{2}\\
     \end{array}\right)
     =
     \left(\begin{array}{c}
     -2 b_1\varepsilon_{xx}\sin\theta_0\cos\theta_0 \\
     0\\
     \end{array}\right)\label{eq:driving_OOP_eps_xx},
\end{equation}
\begin{equation}
\mu_0\left(\begin{array}{c}
     h_{1} \\
     h_{2}\\
     \end{array}\right)
     =
     \left(\begin{array}{c}
     0 \\
     -2b_2\varepsilon_{xy}\sin\theta_0\\
     \end{array}\right)\label{eq:driving_OOP_eps_xy},
\end{equation}
and
\begin{equation}
\mu_0\left(\begin{array}{c}
     h_{1} \\
     h_{2}\\
     \end{array}\right)
     =
     \left(\begin{array}{c}
     -2b_2\varepsilon_{xz}\cos(2\theta_0) \\
     0
     \end{array}\right).\label{eq:driving_OOP_eps_xz}
\end{equation}
For the other out-of-plane configuration investigated (i.e.~$\phi_0=\pi/4$), referred to as OOP2 configuration, the driving fields read as
\begin{equation}
\mu_0\left(\begin{array}{c}
     h_{1} \\
     h_{2}\\
     \end{array}\right)
     =
     \left(\begin{array}{c}
     -b_1\varepsilon_{xx}\sin\theta_0\cos\theta_0 \\
     b_1\varepsilon_{xx}\sin\theta_0\\
     \end{array}\right)\label{eq:driving_OOP_2_eps_xx},
\end{equation}
\begin{equation}
\mu_0\left(\begin{array}{c}
     h_{1} \\
     h_{2}\\
     \end{array}\right)
     =
     \left(\begin{array}{c}
     -b_2\sin\theta_0\cos\theta_0\varepsilon_{xy} \\
     0
     \end{array}\right),\label{eq:driving_OOP_2_eps_xy}
\end{equation}
and
\begin{equation}
\mu_0\left(\begin{array}{c}
     h_{1} \\
     h_{2}\\
     \end{array}\right)
     =
     \left(\begin{array}{c}
     -\sqrt{2}b_2\varepsilon_{xz}\cos(2\theta_0) \\
     \sqrt{2}b_2\varepsilon_{xz}\cos\theta_0\\
     \end{array}\right).\label{eq:driving_OOP_2_eps_xz}
\end{equation}

\end{document}